\documentclass{elsart}

\usepackage{graphicx}

\usepackage{amssymb}

\newcommand{\Sec}[1]{Sec.~\ref{#1}}

\newcommand{\Fig}[1]{Fig.~\ref{#1}}

\newcommand{\Tab}[1]{Tab.~\ref{#1}}
\newcommand{\Eq}[1]{Eq.~(\ref{#1})}

\newcommand{\mytheta}{\theta}
\newcommand{\cms}{c.m.}
\newcommand{\gcm}{\ensuremath{\gamma_{\rm \cms}}}
\newcommand{\Thlabi}[1]{\ensuremath{\mytheta_{\rm lab,#1}}}
\newcommand{\Thcms}{\ensuremath{\mytheta_{\rm \cms}}\ }

\newcommand{\grad}{\ensuremath{^{\circ}}}

\newcommand{\chiz}{\ensuremath{{\chi^2}}}
\newcommand{\eN}{\ensuremath{e_0}}

\newcommand{\tC}{\ensuremath{t_c}}
\newcommand{\NC}{\ensuremath{N_c}}
\newcommand{\Arg}{\ensuremath{(x,\tC)}}
\newcommand{\Argc}{\ensuremath{(x,\tC, c)}}

\newcommand{\rhoHb}{\ensuremath{\overline{\varrho_N}}}

\newcommand{\rhoHN}{\ensuremath{\varrho_N}}
\newcommand{\CH}{\ensuremath{{\rm CH}_2}}
\newcommand{\rch}{\ensuremath{\varrho_{\rm CH_2}}}
\newcommand{\Ach}{\ensuremath{M_{\rm CH_2}}}
\newcommand{\Gy}{{\rm Gy}}
\newcommand{\Tcut}{\ensuremath{T_{\rm cut}}}
\newcommand{\REL}{\ensuremath{\left.\frac{{\rm d}E}{{\rm
d}z}\right|_{T<T_{\rm cut}}}}
\newcommand{\sx}{\ensuremath{\sigma_x}}
\newcommand{\sy}{\ensuremath{\sigma_y}}
\newcommand{\spi}{\ensuremath{\sqrt{2\pi}}}
\newcommand{\Npp}{\ensuremath{N_{\rm pp}}}
\newcommand{\NpC}{\ensuremath{N_{\rm pC}}}
\newcommand{\spp}{\ensuremath{\sigma_{\rm pp}}}
\newcommand{\dd}[1]{\ensuremath{{\rm d}#1}}
\newcommand{\dL}{\dd{L}}

\newcommand{\dtC}{\dd{\tC}}
\newcommand{\dx}{\dd{x}}
\newcommand{\dD}{\dd{D}}

\newcommand{\nh}{\ensuremath{n_{\rm H}}}
\newcommand{\nhn}{\ensuremath{n_{\rm H,0}}}
\newcommand{\dsp}{\displaystyle}

\begin{document}

\begin{frontmatter}


\title{Radiation Damage of Polypropylene Fiber Targets in Storage
Rings\thanksref{support}}
\thanks[support]{Supported by the BMBF and FZ J\"ulich}


\author[ISKP]{ H.~Rohdje\ss\corauthref{cor1}}
\ead{rohdjess@iskp.uni-bonn.de}
\author[HH]{ D.~Albers} 
\author[ISKP]{ J.~Bisplinghoff} 
\author[HH]{ R.~Bollmann} 
\author[HH]{ K.~B\"u\ss{}er} 
\author[ISKP]{ O.~Diehl} 
\author[HH]{ F.~Dohrmann} 
\author[ISKP]{ H.-P.~Engelhardt} 
\author[ISKP]{ P.~D.~Eversheim} 
\author[HH]{ J.~Greiff} 
\author[HH]{ A.~Gro\ss{}} 
\author[ISKP]{ R.~Gro\ss{}-Hardt} 
\author[ISKP]{ F.~Hinterberger}
\author[HH]{ M.~Igelbrink} 
\author[HH]{ R.~Langkau} 
\author[IKP]{ R.~Maier}
\author[ISKP]{ F.~Mosel} 
\author[HH]{ M.~M\"uller} 
\author[HH]{ M.~M\"unstermann} 
\author[IKP]{ D.~Prasuhn}
\author[IKP]{ P.~von~Rossen}
\author[ISKP]{ H.~Scheid} 
\author[HH]{ N.~Schirm} 
\author[ISKP]{ F.~Schwandt} 
\author[HH]{ W.~Scobel} 
\author[ISKP]{ H.~J.~Trelle} 
\author[HH]{ A.~Wellinghausen} 
\author[ISKP]{ W.~Wiedmann} 
\author[HH]{ K.~Woller} 
\author[ISKP]{ R.~Ziegler} 
\corauth[cor1]{Heiko Rohdjess, FAX: +49-228-73 2505}
\address[ISKP]{Helmholtz-Institut f\"ur Strahlen- und Kernphysik, Universit\"at Bonn,
Germany}
\address[HH]{Institut f\"ur Experimentalphysik, Universit\"at Hamburg,
Germany}
\address[IKP]{Institut f\"ur Kernphysik, Forschungszentrum J\"ulich, Germany}

\begin{abstract}
Thin polypropylene (\CH) fibers have been used for internal
experiments in storage rings as an option for hydrogen targets. 
The change of the hydrogen content due to the radiation dose applied
by the circulating proton beam has been investigated in the range
$1\cdot10^6$ to
$2\cdot10^8$~Gy at beam momenta of 1.5 to 3~GeV/c by comparing the
elastic pp-scattering yield to that from inelastic p-carbon reactions.
It is found that the loss of hydrogen as a function of applied dose
receives contributions from a fast and a slow component.
\end{abstract}

\begin{keyword}
radiation damage \sep \ polypropylene \sep fiber target \sep elastic
proton-proton scattering \sep storage ring
\PACS 25.40.Cm \sep 29.25.Pj \sep 61.80.Jh \sep 61.82.Pv 
\end{keyword}
\end{frontmatter}

\section{Introduction}
\label{intro}
Polymers are widely used as  target materials in nuclear- and
particle-physics experiment, since they provide an effective proton
target due to the large hydrogen content. As an example polypropylene
fiber targets of $4\times5$~$\mu$m$^2$ cross section have been used
for the EDDA-experiment \cite{albers97} at COSY \cite{Maier:1997} to
measure precise excitation 
functions of the unpolarized differential cross section at beam
energies from 0.5 to 2.5~GeV.

For polypropylene the H:C ratio should be 2:1, however, upon
irradiation of the target this ratio is going to decrease due to
so-called cross-linking \cite{Charlesby56,Bradley84} of the \CH-chains, a process in which 
H$_2$-molecules are released when C--C bonds replace two C-H bonds.
Losses of hydrogen or carbon due to nuclear reactions of the beam
protons are negligible in comparison.

In an internal target experiment both the beam position and width
change during acceleration \cite{Altmeier:2004,Rohdjess:2004b}, such that an experiment will sample
different areas of the target at different beam-momenta
\cite{Rohdjess:2004b}. As the
applied dose will also vary with beam position the effective
thickness of the hydrogen target will become a function of beam
momentum, as soon as the area of the beam-target overlap changes.

In this paper we investigate the dependence of the hydrogen loss on
the applied dose by irradiating a \CH\ fiber target under very
controlled conditions. 
The number density, i.e. hydrogen atoms per volume, for polypropylene
is given by (see \Tab{tab:not} for our notation)
\begin{equation}
\nhn = \frac{2\,\rch\,N_A}{\Ach}.
\end{equation}
which will change when the target has been irradiated with some dose $D$.
If we introduce a normalized hydrogen density \rhoHN, it will be a function of
the applied dose, and thus of position $x$ along the fiber and time: 
\begin{equation}
\rhoHN(x,t) = \rhoHN(D(x,t)) = \frac{\nh(D(x,t))}{\nhn}.
\end{equation}
The aim of this paper is to determine $\rhoHN(D)$. We used a stored
proton beam at the COSY-accelerator at J\"ulich on a \CH-fiber target, and
recorded the ratio of the pp-elastic scattering yield to that of
proton-carbon inelastic scattering as a function of time.
This ratio directly determines dose related changes in \rhoHN. The scattering
yield of elastic pp-scattering allows a precise determination of the
luminosity and the COSY beam parameters, like width and
position. Using this information the applied dose $D(x,t)$ can be calculated
with accuracy.
\begin{table}
\begin{tabular}{r@{\ :\ }p{.7\textwidth}}
\hline\hline
$x$ & coordinate along the fiber target.\\
$y$ & coordinate orthogonal to the beam and fiber. \\
$z$ & coordinate along the COSY-beam axis \\
$\tC$ & time during COSY's machine cycle\\
$\NC$ & number of COSY cycles measured \\
$dy, dz$ & thickness of the fiber in the vertical and longitudinal
directions\\
$A = dy\cdot dz = 20\ \mu{\rm m}^2$ & cross section of the fiber\\
$N_A$ & Avogadros number\\
$\rch = 0.905\frac{\rm g}{\rm cm^3}$ & density of polypropylene\\
$\Ach = 14\frac{\rm g}{\rm mol}$ & molar mass of \CH \\
$\sx(\tC), \sy(\tC)$ & horizontal and vertical beam width (standard deviation)\\
$x_b(\tC)$ & horizontal beam position (centroid)\\
$I$   &  beam current. \\
$\eN$   & elementary charge.\\
p     & beam momentum.\\
$\frac{\dd{m}}{\dd{l}}$ & mass per unit length of the fiber.\\
\REL(p) & restricted energy loss \cite{pdg02} of protons in polypropylene.\\
\nh  & number density of hydrogen atoms in polypropylene.\\
$\nhn$  & number density of hydrogen atoms in undamaged polypropylene
	$\nhn = 2 N_A \rch/\Ach$.\\
$\rhoHN = \nh/\nhn$ & density of hydrogen atoms normalized to the
value of undamaged polypropylene.\\
$D(x,t)$ & dose distribution along the fiber target as a function of time.\\
\hline\hline
\end{tabular}
\caption{Notation and variables used throughout this paper.}
\label{tab:not}
\end{table}

In section 2 the EDDA experiment is described, the formalism used to
analyze the data in order to extract the hydrogen density as a
function of dose is the topic of section 3. The results are presented
and discussed in section 4.

\section{Experiment}
\label{experiment}

\begin{figure}
\begin{center}
\includegraphics[width=\textwidth]{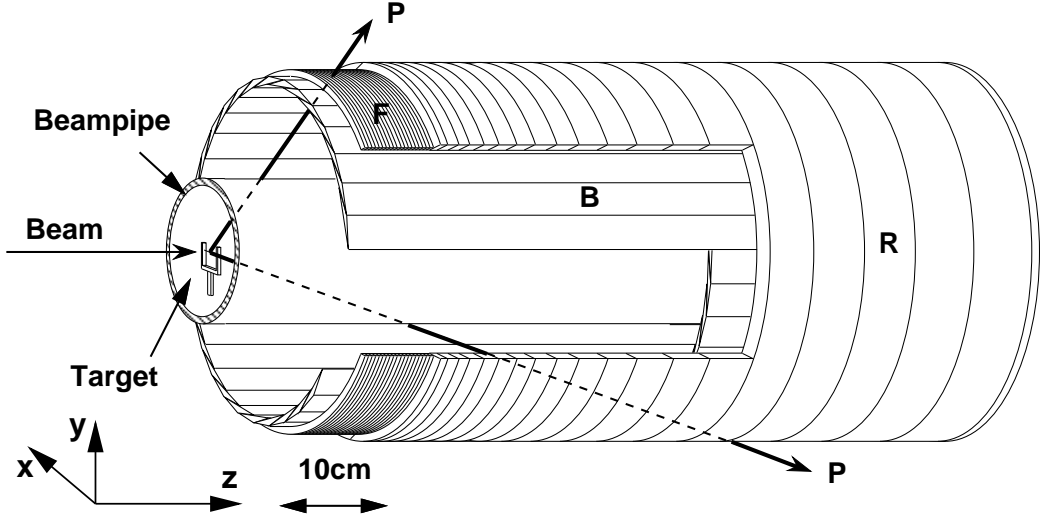}
\end{center}
\caption{ The EDDA detector (not to scale):
          target: fiber (\CH\ or C); 
          B: scintillator bars; R: scintillator semi-rings; 
          F: semi-rings made of scintillating fibers.
	}
\label{fig:detector}
\end{figure}
The EDDA experiment \cite{albers97,Altmeier:2000,Bauer:2002zm} 
has been designed to measure differential cross sections, analyzing
powers and spin correlation parameters of elastic proton-proton
scattering over a wide energy (0.5-2.5~GeV) and angular (30\grad -
90\grad\ in the \cms) range. A schematic view of the EDDA detector is
shown in \Fig{fig:detector}. It is a cylindrical scintillator
hodoscope optimized to efficiently detect two protons from elastic pp
scattering in coincidence. Elastic events are identified by
coplanarity with the beam and the kinematic correlation of the
scattering angles $\tan\Thlabi{1}\cdot\tan\Thlabi{2} = \gcm^{-2}$,
where \gcm\ is the Lorentz-parameter of the c.m. system in the lab.

For this study scattering data acquired at fixed
energies of 3.0 and 1.455 GeV/c with \CH\ and carbon targets, detected
primarily for the purpose of detector calibration, have been used.
The fibers are suspended to a 30~mm wide fork and coated with
20~$\mu$g/cm$^2$ aluminum to avoid charge buildup on the target.
The targets are mounted on a linear actuator and can be moved in and
out of the beam.
The experiment is conducted during many COSY machine cycles: after
beam injection into COSY, the beam is accelerated to the desired beam
momentum, then the fiber target is moved in and data is taken. Since
beam-lifetimes are of the order of a few seconds the target is moved
out after 5 to 10 seconds, and COSY is reset to be prepared for the
next cycle. 

\section{Analysis}
\label{analysis}

For the analysis data has to be sorted as a function of the time
\tC\ an event occurred within a cycle, i.e. \tC\ is the time elapsed
since moving in the target. Due to beam-heating by the target, the width and
eventually the position of the COSY beam, and thus the dose
distribution, change with \tC, however, 
they are stable from one cycle to another.
The target was first irradiated for about 5~h at a beam momentum of
3.0~GeV/c and then for another 5~h at 1.455~GeV/c. 
The complete data set was subdivided in small sets comprising data of
typically 70 machine cycles, corresponding to roughly 15 minutes of
data taking.

\subsection{Overview}

The dose per time interval \dtC\ acquired over \NC\ machine cycles is
a function of the horizontal position $x$ along the fiber and the time
\tC\ during COSY's machine cycle.
It is given by the product
of the average energy deposit $\Delta E$ per particle and the number of
particles hitting the target per unit length of
the fiber divided by its mass per unit length:
\begin{equation} \label{eq:dose1}
\frac{\dD}{\dtC}\Arg = 
\overbrace{\REL\!\!\!\!\!\!\!\!\!\!\!\!\!\!\!(p) \ dz}^{\Delta E}
\
(\dsp\sum_{c=1}^{\NC}\overbrace{\frac{I(\tC,c)\,dy}{\eN\,\spi\,\sy}}^{\d{N_p}/\dtC})
\ \overbrace{\frac{1}{\spi\,\sx} e^{-\frac{(x-x_b)^2}{2\sx^2}}}^{\rm
beam\ profile}
\ \overbrace{\frac{1}{\rch\,A}}^{{\rm d}l/{\rm d}m}.
\end{equation}
Of these quantities the beam current $I$ and the vertical beam width
$\sy$ are hard to measure precisely. However, the luminosity --~with
respect to hydrogen~-- along the target contains the same parameters
\begin{equation}\label{eq:lumi}
\frac{\dL_H}{\dx}\Arg = \sum_{c=1}^{\NC} \overbrace{\frac{I(\tC,c)\,dy}{\eN\,
\spi\,\sy}}^{\d{N_p}/\dtC}  
\ \overbrace{\nhn\,\rhoHN\Argc\,dz}^{\rm target\ area\ density}\
\overbrace{\frac{1}{\spi\,\sx} e^{-\frac{(x-x_b)^2}{2\sx^2}}}^{\rm
beam\ profile}.
\end{equation}
In the experiment we access the luminosity by detecting 
elastic scattering events \Npp\ for \NC\
machine cycles as a function of \tC. Due to the finite resolution in 
determining the x-coordinate of the scattering vertex only the total
luminosity, e.g. integrated over $x$, is available:
\begin{equation}
L_H(\tC,\NC) 
= \dsp \int_{-\infty}^{\infty}\dx \frac{\dL_H}{\dx}\Arg
= \dsp \frac{\Npp(\tC)}{\spp \,[1-\tau(\tC)]\,\epsilon}.
\label{eq:lumiCount}
\end{equation}

Here, \spp\ is the differential cross section integrated over the
accepted solid angle, $\tau$ the deadtime of the data acquisition
system and $\epsilon$ the detection efficiency, i.e. mainly a correction for
losses of elastic events due to secondary reactions in the detector.

Note, that the target density itself changes with the dose applied and will
be both position and time dependent, so that the integration of
\Eq{eq:lumi} cannot be carried out directly. However, we can adopt an
iterative approach and set \rhoHN\ to unity as a first approximation,
which will overestimate the luminosity. The true luminosity will be
smaller by the weighted average over $x$ of the normalized hydrogen density,
introduced via

\begin{equation}\label{eq:rhoAv}
\rhoHb(\tC) =  \frac{\dsp\sum_{c=1}^{\NC}I(\tC,c)\int_{-\infty}^{\infty}\dx \rhoHN\Argc\
\frac{1}{\spi\,\sx} e^{-\frac{(x-x_b)^2}{2\sx^2}}}{\dsp \sum_{c=1}^{\NC}I(\tC,c)},
\end{equation}
a correction, we will have to determine using the relation $\rhoHN(D)$
to be established in \Sec{results}, where in turn $D$ is a function of
$x$, \tC\ and the individual cycle. Again in a first approximation
$\rhoHb$ will be set to one, and recalculated in the following iterations. 
Now we may write \Eq{eq:lumiCount} by inserting \Eq{eq:lumi} as

\begin{equation}
L_H(\tC,\NC) = \rhoHb(\tC) \sum_{c=1}^{\NC} \frac{I(\tC,c)}{\eN}
\cdot\frac{dy\,dz\,\nhn}{\spi\,\sy}
\end{equation}

and thus
 
\begin{equation}
\frac{dy\,dz\,\sum_{c=1}^{\NC} I(\tC,c)}{\spi\,\sy\,\eN} = 
\frac{L_H(\tC,\NC)}{\nhn\,\rhoHb(\tC)} =
\frac{\Ach}{2\,N_A\,\rch}\cdot
\frac{L_H(\tC,\NC)}{\rhoHb(\tC)}.
\end{equation}

Note, that this derivation holds, even when the vertical target
position $y_t$ is not centered at the vertical location of the beam
$y_b$, since the additional factor $\exp(-[y_t-y_b]^2/2\sigma_y^2)$ in
Eqs.~(\ref{eq:dose1}) and (\ref{eq:lumi}) cancels.

Insertion in \Eq{eq:dose1} finally replaces $I$ and $\sigma_y$ by
experimentally accessible quantities

\begin{equation} \label{eq:doseFinal}
\frac{\dD}{\dtC}\Arg = 
\overbrace{\REL\!\!\!\!\!\!\!\!\!\!\!\!\!\!\!(p) \ \ \ 
\ \frac{\Ach}{2\,N_A\,\rch^2\, A}}^{k(p)}\cdot
\frac{L_H(\tC,\NC)}{\rhoHb(\tC)}\cdot
\frac{1}{\spi\,\sx} e^{-\frac{(x-x_b)^2}{2\sx^2}}.
\end{equation}
This must be integrated over \tC\ to obtain the dose profile $D(x)$ along
the beam acquired during \NC\ machine cycles.

The restricted energy loss \REL\ \cite{pdg02} is the energy loss due
to ionization, corrected for the energy carried away by $\delta$
electrons with energies $T > \Tcut$ escaping from the target.
By requiring the practical range of electrons \cite{Gledhill73}
to be half of the \CH-target thickness ($\approx$
0.2~mg/cm$^2$) and taking the thin aluminum-coating ($\approx$
20~$\mu$g/cm$^2$) into account we deduce a value of 9~keV for \Tcut.
Using ionization-loss parameters of \cite{jan82} we obtain numerical
values for the constant $k(p)$ of 190.9 (155.0) ${\rm Gy\,mm\,mb}$ at
1.455 (3.0)~GeV/c.  

To summarize, we need to deduce the integrated luminosity $L_H$, the beam
position and width from the scattering data in order to obtain the
dose profile along the fiber. As pointed out earlier,
Eqs.~(\ref{eq:rhoAv}) and (\ref{eq:doseFinal}) can only be solved
iteratively. One starts by assuming \rhoHb(\tC) being unity, and then
arrives at a dose which will be underestimated. From the measured
change in hydrogen density as a function of the dose
(cf. \Eq{eq:parameterization}) \rhoHb(\tC) is recalculated and a
better estimate of the dose is found. After at most two iteration
self-consistency is reached. 

\subsection{Luminosity}

\begin{figure}
\begin{center}
\includegraphics[width=\textwidth]{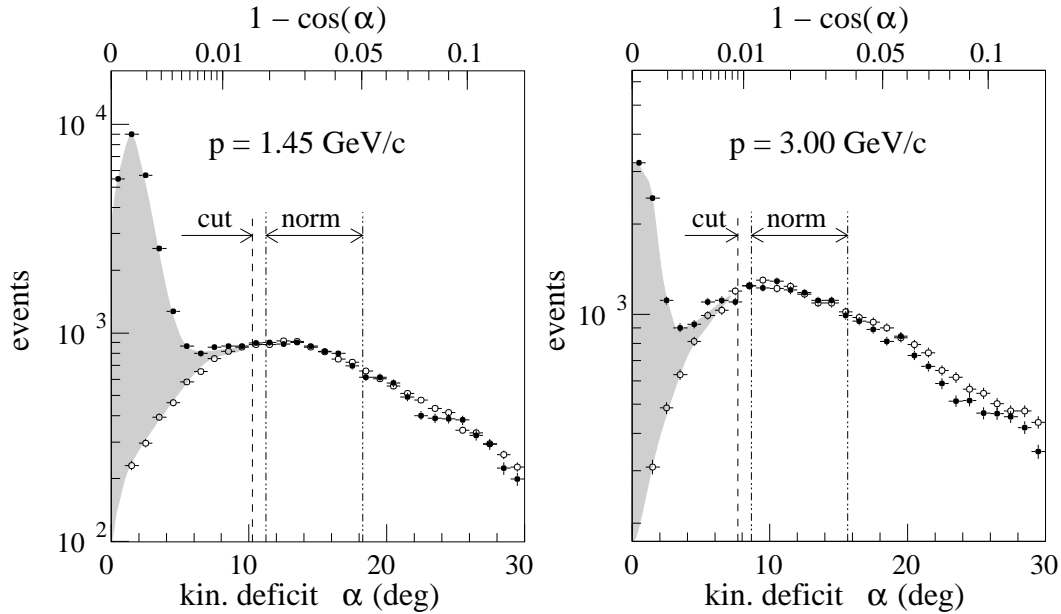}
\end{center}
\caption{Distribution of the kinematic deficit $\alpha$ for data taken with a
\CH (solid) and C-targets (open symbols, normalized to the
\CH-data). The upper limit in $\alpha$ used to select elastic
scattering events (cut) is indicated as well as the range (norm) used
to determine the relative luminosity with respect to scattering of
carbon for the two targets.
}
\label{fig:alpha}
\end{figure}
For the luminosity determination, using \Eq{eq:lumiCount}, we need the
total number of elastic scattering events. These are identified by 
looking at the so-called ``kinematic deficit'', the angle $\alpha$ of the
deviation from a perfect back-to-back correlation of the two detected
prongs in the c.m., when transformed assuming an angle-energy
correlation as for pp elastic scattering. As evinced in
\Fig{fig:alpha}, pp elastic scattering events stand out at very small
$\alpha$ with very little background from proton-carbon scattering. We
use a cut at 10.2\grad (7.7\grad) at 1.455 (3.0)~GeV/c for accepted
events. The geometric acceptance of the EDDA-detector is taken into
account by using only events with polar angles between 36\grad (38\grad) and
90\grad\ in the c.m. for the forward proton. In addition,
events with an azimuthal angle within $\pm15$\grad\ around 90\grad\ or 270\grad\
are excluded, where the angle measurement of \Thcms\ has reduced accuracy
owing to the inhomogeneous light-collection close to the attached light-guides.
The contribution from proton-carbon scattering is removed by subtracting
the yield from scattering of a carbon target using the same cuts and
normalized to the same number of counts in the tails of the
$\alpha$-distribution as indicated in \Fig{fig:alpha}. 

The pp-elastic scattering cross section, cf. \Eq{eq:lumiCount}, is obtained by numerical
integration of the differential cross section as given by the solution
FA00 \cite{Arndt:2000xc} of the VPI/GW phase shift analysis.

Further corrections are applied to account for losses due to secondary
reactions of the ejectiles in the detector (5\%) and for the deadtime
of the data-acquisition event-builder, deduced from the ratio of
triggers read out by the DAQ-system to the total number of triggers,
which is accomplished by 20~MHz scalers, read out every 2.5 ms. Details
are given in \cite{albers97,Altmeier:2004}.
The deadtime ranges from 90\% at the beginning to 10\% at the end of the
cycle. The resulting luminosity is shown at the top of \Fig{fig:vertex}.
 
\subsection{Beam Parameters}
\begin{figure}
\begin{center}
\includegraphics[width=\textwidth]{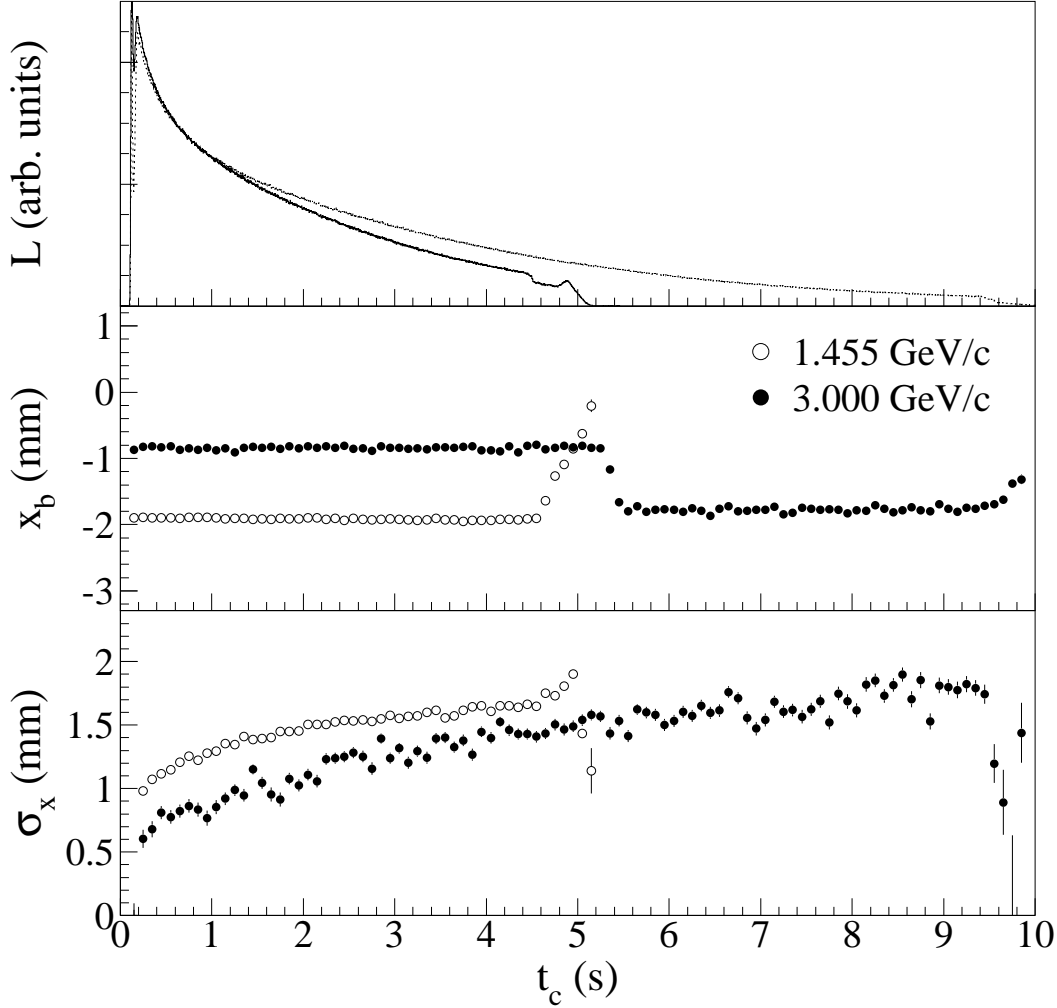}
\end{center}
\caption{Luminosity (top), horizontal beam position (center) and width
(bottom) as a function of time during the COSY-cycle (here \tC=0 has
been set to the time the target is moved in) for beam momenta of 1.455
and 3.3 GeV/c.
}
\label{fig:vertex}
\end{figure}

The beam position and width can be inferred from pp elastic scattering
data as well, by exploiting the coplanarity of the two protons with the
beam. The beam position and width as well as the azimuthal angular
resolution are obtained from a nonlinear \chiz-fit of data with
different values of the azimuthal angle $\phi$. The procedure is
described in detail elsewhere \cite{Rohdjess:2004b}, the results for
the two beam-momenta are displayed in \Fig{fig:vertex}.

The increase of the beam-width or emittance due to the heating of the
stored beam is clearly visible. At 1.455 GeV/c the movement of the
beam at the end of the cycle is due to the beginning deceleration,
whereas the jump in position at mid-cycle of the 3.0~GeV/c data is
caused by exciting 
horizontal steerer used to cross-check the vertex-reconstruction.


\subsection{Hydrogen Density}

The relative change of the hydrogen density in the target is
proportional to the ratio of \Npp, i.e. the shaded area in
\Fig{fig:alpha}, and the number of proton-carbon scattering events
\NpC. For the latter we use the number of events in the range
$11\grad < \alpha < 25\grad$ of \Fig{fig:alpha}, mainly populated by
quasi-free proton-proton scattering. To this end the normalized hydrogen density
is given by 
\begin{equation}
\rhoHN = f(p) \frac{\Npp}{\NpC}
\label{eq:rhoFinal}
\end{equation}
with a momentum-dependent proportionality constant $f(p)$ which needs
to be determined from the data.

\section{Results and Discussion}\label{results}

The data has been subdivided into 43 subsets, 21 at 3.0~GeV/c and 22
at 1.455~GeV/c, and a value for $\rhoHN/f(p)$
has been deduced viz \Eq{eq:rhoFinal}. 
This needs to be related to the dose acquired prior and during data taking of
this subset. Since \rhoHN\ is averaged over the horizontal
beam profile, we relate it to the corresponding average dose
\begin{equation}
\bar{D} = \frac{\dsp\int\dtC\ \Npp(\tC)\ \int\dx\ D(x,\tC)\ \frac{1}{\spi\,\sx}
e^{-\frac{(x-x_b)^2}{2\sx^2}}}{\dsp\int\dtC\ \Npp(\tC)},
\end{equation}
where we implicitly assume, that an appropriate summation over the
number of cycles has been done.
For simplicity, we will denote this dose by $D$ subsequently.

\begin{figure}
\includegraphics[width=\textwidth]{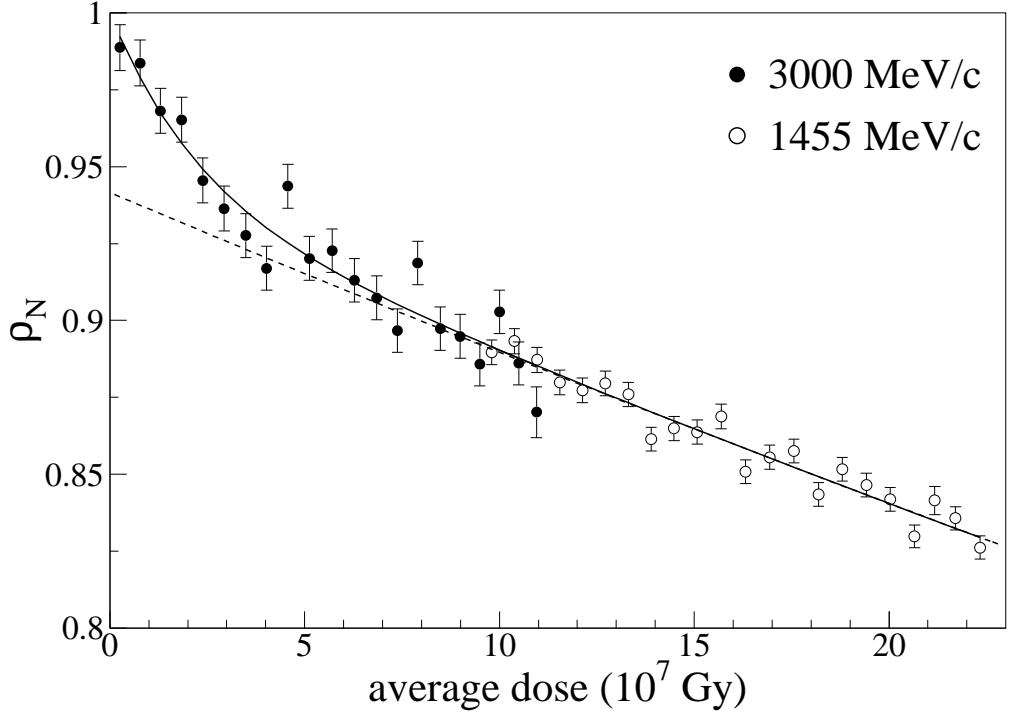}
\caption{Normalized hydrogen density as a function of applied
dose. The result of the fit is shown as the solid line. The
dashed line shows the contribution of the slow component only.
Note, that the first data point at 1.455~GeV/c corresponds to a lower
dose than the last point at 3.0~GeV/c, since the beam is slightly
displaced, such that a region of lower accumulated dose is probed.}
\label{fig:final}
\end{figure}

For the functional dependence of $\rhoHN(D)$ we tried a simple ansatz
of a linear or exponential decrease, which turned out to be
insufficient to describe the data. We obtained a successful fit 
with the following expression: 
\begin{equation}
\rhoHN(D) = \frac{\nh}{\nhn}(D) = a \exp(-\lambda_1 D) + (1 - a) \exp(-\lambda_2 D),
\label{eq:parameterization}
\end{equation}
using a total of five fit parameters $a$, $\lambda_{1,2}$, $f$(1.455~GeV/c), and
$f$(3.0~GeV/c).
The result is shown in \Fig{fig:final}. 

When iterating Eqs.~(\ref{eq:rhoAv}) and
(\ref{eq:doseFinal}), the average hydrogen density for a sample is
given by
\begin{equation}
\overline{\rhoHN} =\dsp \frac{1}{\Delta D}
\int_{D_0}^{D_0+\Delta D} \dD\,\rhoHN(D),
\label{eq:rhobar}
\end{equation}
where $D_0$ and $\Delta D$ are the dose applied prior and during the
measurement of the sample.
After two iterations self-consistency is reached and we obtain the fit values
\begin{equation}
\begin{array}{r@{\ =\ (}l@{\ \pm\ }l@{)\ }l}
\lambda_1 & 4.4 & 1.3 & \cdot 10^{-8}\ \Gy^{-1} \\ 	
\lambda_2 & 5.67 & 0.28 & \cdot 10^{-10}\ \Gy^{-1} \\ 	
        a & 5.8 & 0.7 & \% \\
 \end{array}
\end{equation}
where errors are purely statistical. The systematic error is dominated
by uncertainties in the area $A$ of the target cross-section (10\%) and
the restricted energy loss (4\%), where the exact value depends on the
choice of \Tcut. This leads to a 11\% scale uncertainty of the dose
and consequently of $\lambda_{1}$ and $\lambda_{2}$.

The normalized hydrogen
density \rhoHN\ of \Eq{eq:rhoFinal}, determined from the experiment,
is in fact the average over the length of the fiber.
It receives contributions from regions with quite different
accumulated dose, which are weighted with the beam intensity
distribution and averaged over the machine cycle. It can be shown,
however, that the correct functional dependence is obtained when
\rhoHN(D) is a linear 
function. The systematic effect caused by the deviation from linearity
in \Eq{eq:parameterization} was checked numerically and is always
smaller than 0.25\% in \rhoHN,  and can therefore be neglected.

\begin{figure}
\includegraphics[width=\textwidth]{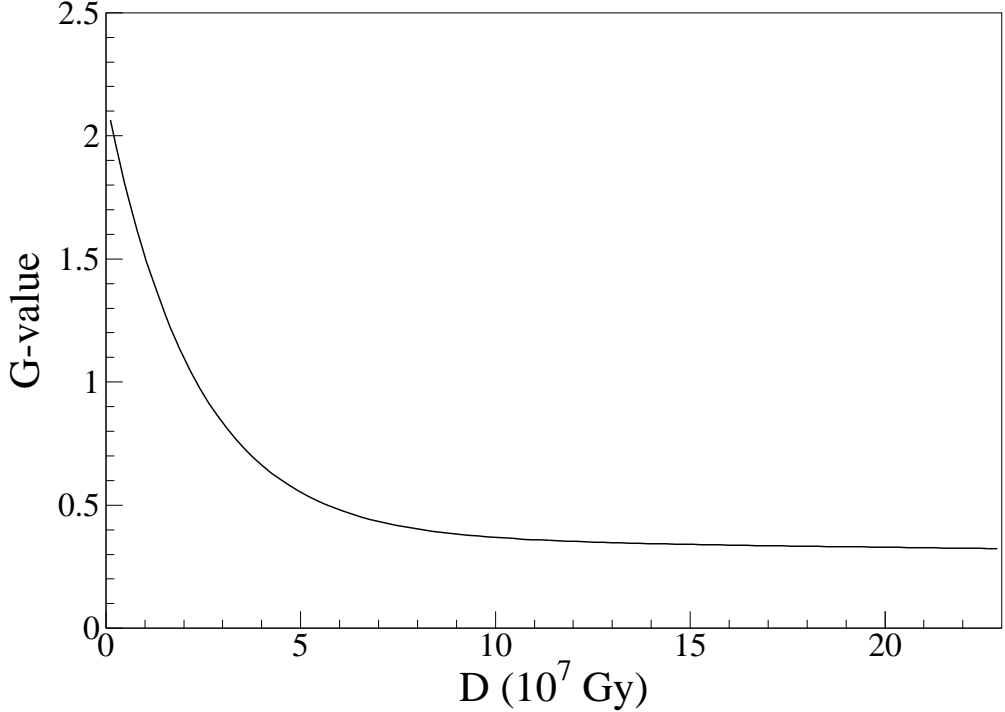}
\caption{G-value for cross-linking as a function of the accumulated dose.}
\label{fig:slope}
\end{figure}

Assuming that the loss of hydrogen is entirely due to cross-linking,
our results may be compared to the so-called $G$-value \cite{Bradley84}
found in the literature. It describes the number of cross links
$\delta N_X$ per $\delta E = 100$~eV energy deposited in the polymer.
Since per cross-link two hydrogen atoms are lost, the change of
hydrogen density for a sample of volume $V$ and mass $m$ is given by
\begin{equation} 
- \frac{\dd{\rhoHN}}{\dd{D}} \approx 
\frac{1}{\nhn}\cdot \frac{2\delta N_X}{V} \cdot \frac{m}{\delta E} 
=
\frac{\rch}{\nhn} \cdot\frac{2
G}{\delta E} = \frac{\Ach}{N_A} \cdot \frac{G}{\delta E}.
\end{equation}
Solving for G using \Eq{eq:parameterization} yields
\begin{equation} 
G = \frac{\delta E\  N_A}{\Ach}
\left[ a \lambda_1 \exp(-\lambda_1 D) + \lambda_2 (1 - a) \exp(-\lambda_2 D)\right]
\end{equation}
which is displayed in \Fig{fig:slope}. It nicely agrees with the range 
of G-values 0.3 to 1.1 \cite{Wood94} and $0.6\pm0.1 $\cite{mosel94}
found in the literature.  However, we find that the G-value is not a 
constant but depends on the irradiation-history of the target. 
The data of \cite{mosel94,mosel91}, obtained by a similar measurement with
14.5~MeV protons on 14~$\mu$m thick polypropylene fibers but with
much lower statistical precision, are in agreement with this finding.

\section{Summary}
Thin \CH-fiber targets have been irradiated by 1.455 and 3.0~GeV/c
protons at an internal target station of the COSY accelerator. The
applied dose distribution was deduced from  the elastic scattering
rate measured continuously during irradiation. The hydrogen density
was monitored by observing the relative yield of proton-proton elastic
scattering with respect to proton-carbon inelastic reactions.
The hydrogen loss as a function of dose can be described by a fast
component responsible for at most 6\% of the total loss, and a slow
component corresponding to G-value of about 0.35. 
The loss rate, attributed to cross-linking of the chains within the
polymer, is in-line with published values. Detailed knowledge
of the observed strong dependence on the accumulated dose of a
polypropylene-target will be important for experiments aiming at
absolute cross-sections, where the hydrogen-density needs to be known
accurately as a function of time as well as vertex position that may
vary considerably during acceleration \cite{Altmeier:2004}.

\section*{Acknowledgments}
We thank the operating team of COSY for excellent beam 
support. This work was supported by the BMBF and Forschungszentrum
J\"ulich GmbH.


\end{document}